\documentclass[10pt,conference]{IEEEtran}
\IEEEoverridecommandlockouts
\usepackage{cite}
\usepackage{amsmath,amssymb,amsfonts}
\usepackage{algorithmic}
\usepackage{graphicx}
\usepackage{textcomp}
\usepackage{booktabs}
\usepackage{rotating}
\usepackage{caption}
\usepackage{multirow}

\usepackage[most]{tcolorbox}
\usepackage{url}%
\usepackage{hyperref}
\usepackage{blindtext}
\usepackage{pifont}
\usepackage{todonotes}

\usepackage{xcolor}
\def\BibTeX{{\rm B\kern-.05em{\sc i\kern-.025em b}\kern-.08em
    T\kern-.1667em\lower.7ex\hbox{E}\kern-.125emX}}
\begin{document}

\title{AutoMESC: Automatic Framework for Mining and Classifying Ethereum Smart Contract Vulnerabilities and Their Fixes}

\author{\IEEEauthorblockN{Majd Soud\IEEEauthorrefmark{1},
Ilham Qasse\IEEEauthorrefmark{1}, Grischa Liebel\IEEEauthorrefmark{1}
Mohammad Hamdaqa\IEEEauthorrefmark{1} \,\IEEEauthorrefmark{2}
}

\IEEEauthorblockA{\IEEEauthorrefmark{1}Department of Computer Science,
Reykjavik University, Reykjavik, Iceland \\
\IEEEauthorrefmark{2}Department of Computer and Software Engineering, Polytechnique Montreal, Montreal, Canada \\
\IEEEauthorrefmark{1}majd18@ru.is, ilham20@ru.is, grischal@ru.is, mhamdaqa@ru.is, \IEEEauthorrefmark{2}mhamdaqa@polymtl.ca
}
}

\maketitle

\begin{abstract}
Due to the risks associated with vulnerabilities in smart contracts, their security has gained significant attention in recent years. 
However, there is a lack of open datasets on smart contract vulnerabilities and their fixes that allows for data-driven research.
Towards this end, we propose an automated method for mining and classifying Ethereum's smart contract vulnerabilities and their corresponding fixes from GitHub and from the Common Vulnerabilities and Exposures (CVE) records in the National Vulnerability Database.
We implemented the proposed method in a fully automated framework, which we call AutoMESC. AutoMESC uses seven of the most well-known smart contract security tools to classify and label the collected vulnerabilities based on vulnerability types. Furthermore, it collects metadata that can be used in data-intensive smart contract security research (e.g., vulnerability detection, vulnerability classification, severity prediction, and automated repair).
We used AutoMESC to construct a sample dataset and made it publicly available. Currently, the dataset contains 6.7K smart contracts' vulnerability-fix pairs written in Solidity.
We assess the quality of the constructed dataset in terms of accuracy, provenance, and relevance, and compare it with existing datasets.
AutoMESC is designed to collect data continuously and keep the corresponding dataset up-to-date with newly discovered smart contract vulnerabilities and their fixes from GitHub and CVE records. 
\end{abstract}


\section{Introduction}
\label{sec:01_introduction}
Smart contracts are computerized self-executing contracts that contain clauses that are enforced once certain conditions are met \cite{kolvart2016smart}. The concept of smart contract was first introduced in 1996 by Szabo~\cite{szabo1996smart}. 
However, due to technology limitations at that time, the first real-world implementation of a smart contract was developed by the Ethereum blockchain platform in 2014 \cite{buterin2014next}.

The evolution of smart contracts has enabled blockchain technology to grow rapidly and has led to the adoption of decentralized applications in various fields such as the Internet of Things (IoT), supply chain management, and identity management \cite{zheng2020overview}. 
While smart contracts have the potential to reshape how businesses are carried out, several challenges need to be addressed \cite{zheng2020overview} that are not commonly found in traditional software development \cite{marchesi2020design,vacca2021systematic}. 
For instance, immutability of smart contracts \cite{zhang2020framework} can lead to financial losses, such as the infamous DAO attack that led to a theft of approximately 50 million US dollars \cite{mehar2019understanding}. 

To improve the security of smart contracts, several methods and tools have been suggested to detect and fix security vulnerabilities in smart contract source code \cite{rameder2021systematic}. 
For evaluation, these contributions rely on existing datasets \cite{bhandari2021cvefixes,choi2017end}.
However, existing datasets, e.g., \cite{sujeet2022scrawld,durieux2020empirical,zhang2020framework,ren2021empirical}, are limited in one of the following ways: 
(i) collecting and labeling the data is time and resource-intensive, e.g., \cite{sujeet2022scrawld}, (ii) they are limited in terms of the amount of metadata they contain, e.g., \cite{durieux2020empirical}, (iii) they are incompletely labeled or classified, e.g., \cite{SmartEmbed}, (iv) fixes of the labeled vulnerabilities are not considered, e.g, \cite{ren2021empirical}, and (v) they are not updated regularly, over time resulting in invalid or outdated data, e.g. \cite{zhang2020framework}.

To address these limitations, we propose a method to automatically mine, classify and label smart contract vulnerabilities and corresponding fixes from GitHub\footnote{https://github.com/} and CVE\footnote{https://cve.mitre.org/} records. We implement the proposed method in a fully automated framework, AutoMESC, that curates and classifies the collected vulnerabilities based on their CWE and vulnerability types using 7 of the most well-known smart contract security tools in the literature. We target Ethereum smart contracts written in Solidity\footnote{https://docs.soliditylang.org/en/v0.8.13/} and Vyper\footnote{https://vyper.readthedocs.io/en/stable/}, the two most common languages for Ethereum smart contracts.
AutoMESC places an emphasis on extracting a large number of attributes and labels, so that the resulting datasets can be used for use cases that rely on large number of labeled data, e.g., machine learning (ML) applications.


\textbf{Contributions.} To summarize the most salient contributions of our research, we:
\begin{enumerate}
 \item provide a comprehensive survey and comparison of the available datasets for smart contract vulnerabilities written in the two most popular Ethereum languages, Solidity and Vyper. To the best of our knowledge, this is the first comprehensive study of existing Ethereum smart contract vulnerability datasets.  
 \item present a method for mining and classifying smart contract vulnerabilities and corresponding fixes from GitHub\footnote{\url{https://github.com/}} and CVE\footnote{\url{https://cve.mitre.org/}} records. We implement the proposed method in an automated framework (i.e., AutoMESC).
\item present a sample dataset\footnote{\url{https://figshare.com/s/ea43905535cc1302267b}} extracted using AutoMESC. We evaluate the quality of this dataset in terms of accuracy, relevance, and provenance.
\end{enumerate}

The rest of this paper is organized as follows: Section~\ref{sec:survey} presents the related work. Section~\ref{sec:framework} discusses the method of AutoMESC. Section~\ref{sec:framework} presents and explores the AutoMESC dataset. 
Section~\ref{sec:app} describes the applications of AutoMESC and the extracted dataset. In Section~\ref{sec:results}, we evaluate the AutoMESC dataset and compare it to existing datasets. Section~\ref{sec:threats} discusses the potential threats to validity. The paper is concluded in Section~\ref{sec:conclusion}..

\section{Related Work}
\label{sec:survey}
In this section, we survey existing smart contracts vulnerability datasets. Moreover, we present related automated tools for mining repositories and identifying smart contract vulnerabilities. Finally, we discuss the research gaps in the related work.

\subsection{Smart Contracts Vulnerability Datasets}
Numerous smart contract vulnerability related datasets have been developed over the last few years. Each of the proposed datasets are unique and have different strengths and weaknesses. In the following, we analysed the proposed datasets.

Using Google's dataset search engine\footnote{\url{https://datasetsearch.research.google.com/}} and the keywords \emph{``smart contract vulnerabilities"}, \emph{``smart contract fixes"}, and \emph{``smart contracts"}, we were able to identify 95 relevant datasets containing smart contract vulnerabilities and fixes\footnote{At the end of March 2022.}. 
Then, we applied exclusion criteria in two phases: 
In Phase \#1, we considered recent datasets (up to three years)
as smart contract languages are evolving quickly. In addition to limited years, we considered public datasets and excluded non-free and closed datasets. Finally, in Phase \#1, we excluded commercial and competition datasets and included academia-related datasets. After applying the Phase \#1 exclusion criteria, we were left with 45 datasets.
    
In Phase \#2, we excluded any dataset that is not related to Ethereum smart contract vulnerabilities. Moreover, we only considered datasets with published papers/preprints or a public GitHub repository. Finally, we excluded datasets that only contained sample contracts, e.g., one contract for each vulnerability type, as they are not relevant to the scope of this paper. After applying the criteria in Phase \#2, 5 datasets remained. Detailed information about the included and excluded datasets (i.e., the name and URL for each dataset), as well as the reasons for exclusion, has been published publicly \footnote{\url{https://doi.org/10.5281/zenodo.6762730}}. Following are the five studies with the datasets that were included.

Yashavant et al.~\cite{sujeet2022scrawld} constructed a dataset called ScrawlD of real world Ethereum smart contracts labeled with vulnerabilities. It was created to support unbiased evaluation of existing tools that were proposed in the literature for analyzing smart contracts. By majority vote, the data set was labeled by 5 tools that identify various vulnerabilities.
The dataset has 6.7k labeled Ethereum smart contracts extracted from Etherscan~\cite{Etherscan}. The authors acknowledge that the proposed approaches for labeling the contracts (i.e., integrating the available tools or manually labeling the rest of contracts) are time and resource-intensive, therefore the authors plan to do so incrementally and update the dataset regularly in the future. The dataset is available in ScrawlDset~\cite{ScrawlDset}.  

Durieux et al.~\cite{durieux2020empirical} presented two novel datasets with the goal of evaluating the precision of state-of-the-art smart contract analysis tools. The first dataset consists of 69 annotated vulnerable smart contracts and the second consists of 47,518 contracts extracted from Etherscan~\cite{Etherscan}. The first dataset is curated and labeled with the location and category of the vulnerabilities, while the second dataset consists only of smart contracts written in Solidity, and the vulnerabilities of those contracts is unknown. In our paper, we consider datasets with vulnerabilities and/or fixes, so we only include the first dataset. The first dataset was labeled using the DASP10\footnote{\url{https://dasp.co/}} taxonomy of Ethereum smart contracts vulnerabilities. The labeling of the first dataset is on two code levels: the vulnerability type and the line numbers where the vulnerability occurs in the contract. 
Labeling the 69 vulnerable contracts resulted in 115 labeled vulnerabilities, divided into 10 vulnerability types. 
The dataset is available on \cite{smartbug}.

Ren et al.~\cite{ren2021empirical} constructed a dataset with 46,186 diversified contracts crawled from Etherscan, SolidiFI repository, CVE and Smart Contract Weakness Classification and Test Cases library. 
The collected contracts can be classified into three categories: 1) unlabeled real-world contracts; 2) contracts with manually injected bugs; 3) confirmed vulnerable contracts. The dataset is available on GitHub\footnote{\url{https://github.com/renardbebe/Smart-Contract-Benchmark-Suites}}. The labeled dataset has a total number of 350 artificially constructed contracts and 214 confirmed vulnerable contracts.

Zhang et al. \cite{zhang2020framework} constructed a dataset called Jiuzhou. It consists of 176 smart contracts with vulnerabilities. It provides two contracts for every type of the studied vulnerabilities: one with the vulnerability and one without it. 
The authors suggest three applications for the Jiuzhou dataset: (1) developers can learn about smart contract vulnerabilities when reading the labeled contracts, (2) it can guide developers to implement smart contract tools, and (3) it can be used to evaluate the existing tools. 

Gigahorse benchmarks\footnote{\url{https://github.com/nevillegrech/gigahorse-benchmarks}} consists of a collection of Ethereum smart contracts in source and binary format, labeled with respective vulnerabilities. Some contracts in this dataset have been derived from Smartbugs~\cite{durieux2020empirical}. Gigahorse has three main data types: Invulnerable bytecode \footnote{compiled contract}, vulnerable bytecode and vulnerable source. This dataset is still under construction. The dataset also only covers the 10 vulnerabilities defined in the DASP taxonomy.


\subsection{Automated Tools}
This section presents an overview of the related automated tools to AutoMESC including general purpose tools for mining vulnerabilities. 

Ferreira et al. \cite{ferreira2020smartbugs} introduced SmartBugs, an automated framework to analyze smart contracts written in Solidity based on ten security analysis tools for smart contracts. Nevertheless, the tool only analyses smart contract vulnerabilities and doesn’t collect or construct datasets. 


CVEfixes \cite{bhandari2021cvefixes} is a dataset and an automated collection tool that collects vulnerable code and corresponding fixes from open-source software repositories. CVEfixes supports data collection from multiple programming language repositories. However, the CVEfixes dataset does not contain any vulnerable smart contract codes. In addition, CVEfixes relies solely on CVE to classify code vulnerabilities. This is not enough in the case of smart contracts, as there are very few CVE records for smart contracts. 
Furthermore, smart contracts have unique vulnerabilities such as re-enteracy, integer arithmetic errors, extra gas consumption, etc. These vulnerabilities can not be detected by traditional general-purpose security tools, and require specialized tools in smart contracts to label and classify them. 

In addition to CVEfixes, there are many frameworks to mine software repositories such as \cite{spadini2018pydriller,duenas2018perceval}. However, these frameworks target specific programming languages such as Python, Java, etc. These frameworks do not support smart contract programming languages such as Solidity and Vyper.

Table \ref{tab:comp} compares related automated tools and AutoMESC in terms of data collection, supported smart contract languages, number of used tools, and whether vulnerability fixes are supported or not. AutoMESC and Smart Bugs support smart contract programming languages and apply several security analysis tools to detect vulnerabilities. However, SmartBugs does not automatically collect data; it's only for analyzing smart contract code and detecting vulnerabilities. On the other hand, CVEfixes and AutoMESC support data collection automatically, including vulnerability fixes. However, CVEfixes does not support labeling smart contracts' vulnerabilities. 

\begin{table}[]
\caption{Comparison between related automated tools and AutoMESC}
\label{tab:comp}
\begin{tabular}{c|c|c|c|l}
\cline{2-4}
                                             & SmartBugs & CVEfixes  & AutoMESC  &  \\ \cline{1-4}
\multicolumn{1}{|c|}{Data collection}        & No       & Automatic & Automatic &  \\ \cline{1-4}
\multicolumn{1}{|c|}{\begin{tabular}[c]{@{}c@{}}Supported smart contract \\ programming languages\end{tabular}} & Solidity & N/A & Solidity and Vyper &  \\ \cline{1-4}
\multicolumn{1}{|c|}{\# of used tools}       & 10       & N/A       & 7         &  \\ \cline{1-4}
\multicolumn{1}{|c|}{Volunaribilities fixes} & No       & Yes       & Yes       &  \\ \cline{1-4}
\end{tabular}
\end{table}

\subsection{Research Gaps}
Based on the related datasets and frameworks, we identified the following research gaps: 

 \textbf{Datasets that support data-driven approaches are scarce}.
 Most of the available datasets were never created for the purpose of supporting data-driven research in the field of smart contracts vulnerabilities or fixes. Existing datasets are constructed with the purpose of evaluating state-of-the-art tools. For instance, \cite{sujeet2022scrawld} and \cite{ren2021empirical} aim to eliminate bias in assessing smart contracts security analysis tools. Also, \cite{durieux2020empirical} and \cite{zhang2020framework} support the evaluation of smart contract security analysis tools. \\
Moreover, most available datasets are not labeled or contain a few labeled vulnerable contracts. These are very general datasets, where researchers still need to conduct labeling, analysis, and pre-processing to utilize the data with data-driven models. Therefore, there is a need for datasets with diverse samples of smart contracts vulnerabilities and their fixes for reliable training and evaluation of ML, or deep learning approaches \cite{russell2018automated,coulter2020code,lin2020software}. 
 \\
 \textbf{Existing datasets are not updated regularly.} Several vulnerability types become deprecated over time, and there is a need to collect vulnerabilities that developers encounter on a day-to-day basis. Regularly updating the dataset on newly disclosed vulnerabilities may be a possible solution. \\\
 \textbf{In the labeled datasets, fixes to the labeled vulnerabilities are not addressed.} There is a lack of databases covering smart contract vulnerability fixes or the relationships among them. This limits the data-driven repairing methods and research. In addition, it limits automation and empirical software research on vulnerability-fixes and fixes patterns. \\
\textbf{There is no variation in the level of granularity of labeled datasets}, since most of them are at the contract level. According to \cite{zou2019mu}, levels of granularity and the precise location of a vulnerability in the dataset are finer than the widely used granularity of programs and files. Also, \cite{morrison2015challenges} conclude that file-level granularity decreases precision and recall performance of analysis tools. Thus, multiple granularity levels are needed in the current available datasets to support data-driven research on smart contracts. \\
 \textbf{There is a lack of datasets that supports Vyper Ethereum smart contract language research}. In our review, we only found one dataset written in Vyper and published by vyperhub-io \cite{Vyperhub}. However, by the time of writing this paper there is no publicly available labeled dataset of Vyper smart contract vulnerabilities. This is an important shortcoming, since there are many recently disclosed Vyper related vulnerabilities with high severity at the NVD database e.g. CVE-2022-24845\footnote{\url{https://nvd.nist.gov/vuln/detail/CVE-2022-24845}}, CVE-2022-24788 \footnote{\url{https://nvd.nist.gov/vuln/detail/CVE-2022-24788}}, and CVE-2021-41121\footnote{\url{https://nvd.nist.gov/vuln/detail/CVE-2021-41121}}. 
 
 The AutoMESC framework and dataset presented in this paper address these research gaps in the following ways: (1) It utilizes 7 of the most well-known smart contracts security tools to classify and label smart contracts vulnerabilities written in Solidity or Vyper. These tools support detecting a wider range of smart contract vulnerabilities (36 vulnerability types). (2) It includes vulnerable smart contract codes and the corresponding fixes (6.7K) at different levels of granularity. The dataset includes the vulnerable code lines, the corresponding fixes, and the line changes in the fixed code. (3) It updates regularly every two hours, where newly disclosed vulnerabilities are added. 
This enables AutoMESC dataset to support data-driven research in the field of smart contract vulnerabilities or fixes. 






 


\section{Details of AutoMESC} \label{sec:framework}

\begin{figure*}[htbp]
\centering
\includegraphics[width=0.96\textwidth]{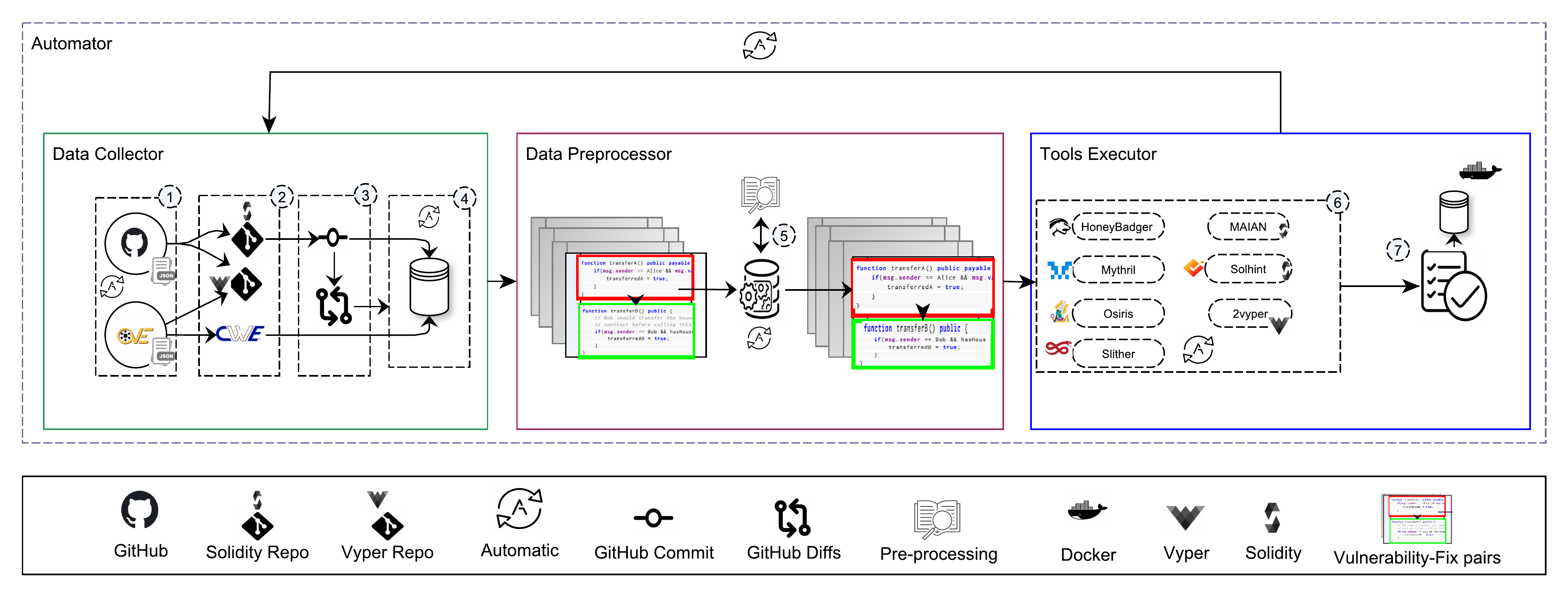}
\caption{AutoMESC Framework high-level architecture}
\label{fig:method}
\end{figure*} 

In this section, we describe the proposed method to mine, classify and label smart contract vulnerabilities and fixes as a fully automated framework, AutoMESC. 
We discuss data collection, data curation, automatic execution of candidate tools, and output. 

Figure \ref{fig:method} illustrates the high-level architecture of AutoMESC, which is composed of three main components: Data Collector, Data Preprocessor, and Tool Executor.
The Data Collector (DC) mines existing repositories to collect data from GitHub and CVE records (i.e., steps 1, 2, and 3) and stores the extracted data in a relational database (i.e., step 4). The Data Preprocessor (DP) pre-processes stored data (i.e., step 5). The Tool Executor (TE) runs seven vulnerability detection tools on the data to label the vulnerabilities in step 6 and finally stores the final data in step 7.
Results are stored in the Database, and the Automator component automates the process of collecting and analyzing data.

%


\subsection{Data Collector Methodology} \label{M:collectdata}
The data flow process of AutoMESC starts with automatic mining and collection of data from open-source software (OSS) projects on GitHub and the National Vulnerability Database (NVD). Using the GitHub API, it automatically collects vulnerability-fix pairs that developers have contributed, written in Solidity or Vyper. It also automatically collects vulnerability records (CVEs) using the JSON vulnerability feed published by the NVD database, organized by the year of origin. In order to keep up to date with newly discovered and patched smart contract vulnerabilities, the automatic collection is continuously repeated from both OSS projects hosted on GitHub and the NVD database.

\subsubsection{Mining OSS projects} 
\begin{figure}[ht]
\centering
\includegraphics[width=0.5\textwidth]{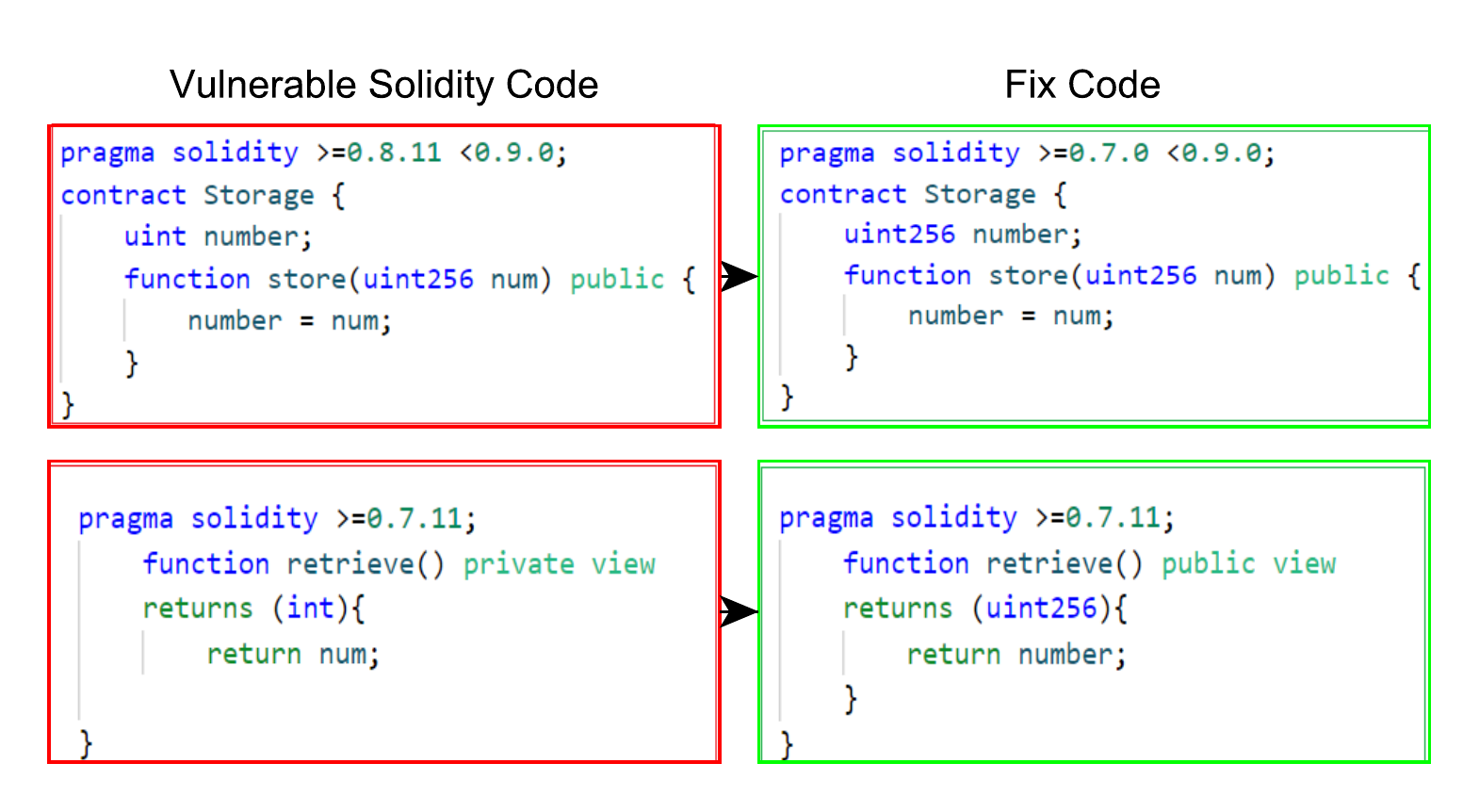}
\caption{Smart contract vulnerability-fix pair example}
\label{fig:vpexample}
\end{figure}

The main source of smart contract vulnerabilities and corresponding fixes in AutoMESC is GitHub. AutoMESC first starts mining all the projects hosted on GitHub that have either Solidity or Vyper as their main language. After that, AutoMESC retrieves all the related metadata for these projects and stores them in the database. AutoMESC collects all commits related to vulnerabilities and fixes, using regular expressions based on selected keywords and related to \emph{``smart contract"}, \emph{``Solidity"}, \emph{``Vyper"} and \emph{``Ethereum"}. The selected keywords, illustrated in Table \ref{tab:keywords}, are based on the common keywords associated with vulnerabilities presented by Bosu et al. \cite{bosu2014identifying}. Furthermore, all commit fixes are collected and then filtered automatically, as described next in the preprocessing methodology. Fixes not related to code are automatically cleaned so that the data can be used for ML models. Vulnerable code is automatically mapped to a corresponding fix code via a hash, resulting in smart contract vulnerability-fix pairs that contain both the code before and after correction. In Figure \ref{fig:vpexample}, we show examples of smart contract vulnerability-fix pairs in the dataset.

AutoMESC also collects the file in which the vulnerability and the corresponding fix occurred, and maps the two files together. The same mapping is done for methods.
Ultimately, the dataset has multiple levels of granularity, i.e., on file, method, and line level.

\begin{table}[]
\caption{Selected keywords for mining OSS projects}
\label{tab:keywords}
\resizebox{\columnwidth}{!}{%
\begin{tabular}{lll}
\cline{1-2}
\multicolumn{1}{|c|}{keywords} &
  \multicolumn{1}{c|}{\begin{tabular}[c]{@{}c@{}}security, vulnerability, vulnerable, exploit, \\ threat, expose, bug, defect, insecure\end{tabular}} &
   \\ \cline{1-2}
 &
   &
   \\
 &
   &
  
\end{tabular}%
}
\end{table}

\subsubsection{Mining CVE Records} AutoMESC automatically collects all published CVEs related to Ethereum smart contracts up until the last published CVE on the date of collection by retrieving the published JSON feeds from the NVD server (i.e., Step 1 in Figure~\ref{fig:method}). It then aggregates and processes the JSON files to remove duplicate CVEs. Afterwards, it collects and stores several details about each vulnerability CVE such as CWE, published date, last modified date, and CVSS severity score (i.e., Step 2 in Figure~\ref{fig:method}). More details about the collected information are listed in Section~\ref{datasetsubsection}.   

\subsubsection{Automation} 
The framework runs in the background and automatically collects newly posted vulnerabilities from OSS projects on GitHub and newly disclosed CVEs on the NVD database every two hours.
The first time AutoMESC is run, it collects all data from 2016\footnote{The first attack on smart contracts was the DAO attack in 2016 \cite{atzei2017survey}.} onwards.

\subsection{Data Preprocessor Methodology}
AutoMESC preprocesses the collected data from GitHub automatically on commit and on code level. On commit level, it excludes any commits that have no code and only considers commits that affect Solidity (.sol) and Vyper (.vy) files.
AutoMESC further analyzes these files (.sol and .vy) and removes comments and white spaces automatically from the files, before storing them in the database.

\subsection{Tool Executor Methodology}
AutoMESC executes vulnerability detection tools on the collected data and labels the data based on the output of the tools. 
\subsubsection{Selected Tools}
In order to label the collected vulnerabilities in AutoMESC, we employ available smart contract state-of-the-art analysis tools for both Solidity and Vyper (as discussed in Section~\ref{sec:survey}). 
We only considered tools that:
\begin{itemize}
    \item detect vulnerabilities or code weaknesses in smart contracts and identify their types. 
    \item take as input either a Solidity (.sol) or Vyper (.vy) source code file.
    \item are publicly available and based on Docker. Docker facilitates the portability of our framework and the scalability of our contract analysis.
\end{itemize}

We selected 7 tools, which we describe in the following.

\textbf{Osiris}~\cite{Osiris} was developed with the goal of detecting integer vulnerabilities in Solidity. It is based on a symbolic execution combined with taint analysis. 
It consists of three core components: symbolic analysis, taint analysis, and integer error detection. 

\textbf{Slither}~\cite{Slither} was designed to statically analyze smart contracts and provide users with information about the analyzed smart contract. 
Slither can automatically detect vulnerabilities via security issue detectors. 

\textbf{SmartCheck} was developed by the SmartDec team \cite{Smartcheck}. It takes Solidity code and detects security vulnerabilities and bad coding patterns. The tool also provide users with the severity levels of the detected vulnerabilities. It also provides warnings for low severity vulnerabilities such as redundant functions.

\textbf{Solhint}~\cite{solhint} is a tool for detecting syntax-related vulnerabilities in smart contracts. It statically analyzes the code and checks a wide range of rules. It also provides developers with the ability to add new rules based on Solidity style guides and manage the configuration rules at the code level. 

\textbf{Honeybadger}~\cite{HONEYBADGER} was designed to detect possible honeypots in smart contracts. A honeypot is a new type of fraud in Ethereum, in which the attacker attracts the victim into traps by deploying vulnerable contracts that contain hidden traps. 

\textbf{Mythril}~\cite{Mythril} allows users to check their smart contracts written in Solidity using solc, a command-line compiler. It can detect several vulnerabilities such as overflow/underflow, and tx.origin. While some severe vulnerabilities were presented as transaction-ordering dependence and information exposure.

\textbf{Maian}~\cite{nikolic2018finding} labels vulnerable smart contracts into three main categories: suicidal contracts, prodigal contracts, and greedy contracts. The tool analyzes Solidity contracts by running a dynamic analysis in a private blockchain, in order to reduce the number of false positives. 

\subsubsection{Vulnerability Labeling}
AutoMESC can detect up to 36 unique vulnerabilities based on the selected tools. However, some tools detect the same vulnerability using different names. We mapped the tools' vulnerabilities to the proposed vulnerability type classification in \cite{soud2022fly} to unify vulnerability labeling. Table \ref{tab:mapping} demonstrates a sample of supported vulnerabilities and the unified label used by AutoMESC. The symbol `` \textbf{-} '' indicates that the tool does not detect the vulnerability. 

\begin{table*}[]
\centering
\caption{Sample of supported vulnerabilities and the different labels used by each supporting tool}
\label{tab:mapping}
\small
\resizebox{\textwidth}{!}{%
\begin{tabular}{|c|c|c|c|c|c|c|c|}
\hline
Vulnerability / Tools &
  Osiris &
  Slither &
  Smart Check &
  Solhint &
  Honeybadger &
  Mythril &
  Maian \\ \hline
Suicidal Contracts &
  - &
  suicidal &
  - &
  avoid-suicide &
  - &
  Suicide &
  suicidal contract \\ \hline
\begin{tabular}[c]{@{}c@{}}Integer Overflow \\ and Underflow\end{tabular} &
  Arithmetic Bugs &
  - &
  Unchecked math &
  - &
  - &
  Integer &
  - \\ \hline
Frozen Ether &
  - &
  locked-ether &
  Locked money &
  - &
  - &
  - &
  Greedy contracts \\ \hline
Reentrancy &
  - &
  \begin{tabular}[c]{@{}c@{}}reentrancy-eth\\ reentrancy-no-eth\\ reentrancy-events\\ reentrancy-unlimited-gas\end{tabular} &
  Reentrancy &
  reentrancy &
  - &
  \begin{tabular}[c]{@{}c@{}}State Change External Calls\\ External Calls\end{tabular} &
  - \\ \hline
Denial of Service &
  - &
  incorrect-equality &
  \begin{tabular}[c]{@{}c@{}}DoS by \\ external contract\end{tabular} &
  multiple-sends &
  - &
  Multiple Sends &
  - \\ \hline
\begin{tabular}[c]{@{}c@{}}Unchecked Call \\ Return Value\end{tabular} &
  - &
  \begin{tabular}[c]{@{}c@{}}unchecked-transfer\\ unchecked-lowlevel\end{tabular} &
  Unchecked external call &
  reason-string &
  - &
  Unchecked Retval &
  - \\ \hline
\begin{tabular}[c]{@{}c@{}}Authorization\\  through tx.origin\end{tabular} &
  - &
  tx-origin &
  Using tx.origin &
  avoid-tx-origin &
  - &
  - &
  - \\ \hline
\begin{tabular}[c]{@{}c@{}}Insecure Contract \\ Upgrading\end{tabular} &
  - &
  unprotected-upgrade &
  - &
  - &
  - &
  \begin{tabular}[c]{@{}c@{}}Delegate Call To \\ Untrusted Contract\end{tabular} &
  - \\ \hline
Gas Costly Loops &
  - &
  costly-loop &
  Costly loop &
  - &
  - &
  - &
  - \\ \hline
Balance Disorder &
  - &
  - &
  - &
  - &
  Balance Disorder &
  - &
  - \\ \hline
\end{tabular}%
}
\end{table*}

Each tool selected for vulnerability detection in AutoMESC is neither sound nor complete. In some cases, a tool may generate a high number of false positives or false negatives. Therefore we can not use one tool output to label the data. Hence, in AutoMESC, we consider the majority rule to label if a vulnerability exists in the data or not. Based on Table \ref{tab:threshold}, at least 50\% of the tools that detect a vulnerability (threshold) must report the same output at the same position.
\begin{table}[t]
\footnotesize
\centering
\caption{Sample of supported vulnerabilities and their threshold}
\label{tab:threshold}
\begin{tabular}{|c|c|c|}
\hline
Vulnerability & \begin{tabular}[c]{@{}c@{}}\# of Tools that can\\  detect the vulnerability\end{tabular} & Threshold \\ \hline
Suicidal Contracts                                                         & 4 & 2 \\ \hline
\begin{tabular}[c]{@{}c@{}}Integer Overflow \\ and Underflow\end{tabular}  & 3 & 2 \\ \hline
Frozen Ether                                                               & 3 & 2 \\ \hline
Reentrancy                                                                 & 4 & 2 \\ \hline
Denial of Service                                                          & 4 & 2 \\ \hline
\begin{tabular}[c]{@{}c@{}}Unchecked Call \\ Return Value\end{tabular}     & 4 & 2 \\ \hline
\begin{tabular}[c]{@{}c@{}}Authorization\\  through tx.origin\end{tabular} & 3 & 2 \\ \hline
\begin{tabular}[c]{@{}c@{}}Insecure Contract \\ Upgrading\end{tabular}     & 2 & 1 \\ \hline
Gas Costly Loops                                                           & 2 & 1 \\ \hline
Balance Disorder                                                           & 1 & 1 \\ \hline
\end{tabular}%
\end{table}

The process of labeling data as a vulnerability includes the following steps:
\begin{enumerate}
    \item Collect the output of all selected tools for each collected data. 
    \item For each output, identify the detected vulnerability name and the location of the vulnerability (line number).
    \item Based on the mapping in Table \ref{tab:mapping}, unify the vulnerability name.
    \item Apply the majority rule to label the collected data. AutoMESC will label a vulnerability if at least 50\% of the tools that detect this vulnerability (based on Table \ref{tab:threshold}) report the same vulnerability at the same position.
\end{enumerate}

AutoMESC uses keyword matching to detect commits fixing vulnerabilities. In practice, this method could generate considerable noise, such as commits not fixing any vulnerabilities but with the selected keywords. To handle these noises, AutoMESC analyses the fixed versions using the selected tools and checks whether the same vulnerability is detected or not. Utilizing the majority rule, if the same vulnerability is found in the fixed code, AutoMESC considers this commit as noise and does not include it in the dataset as a vulnerability-fix pair.



\section{AutoMESC Dataset}
\label{datasetsubsection}
\begin{figure*}[htbp]
\centering
\includegraphics[width=0.8\textwidth]{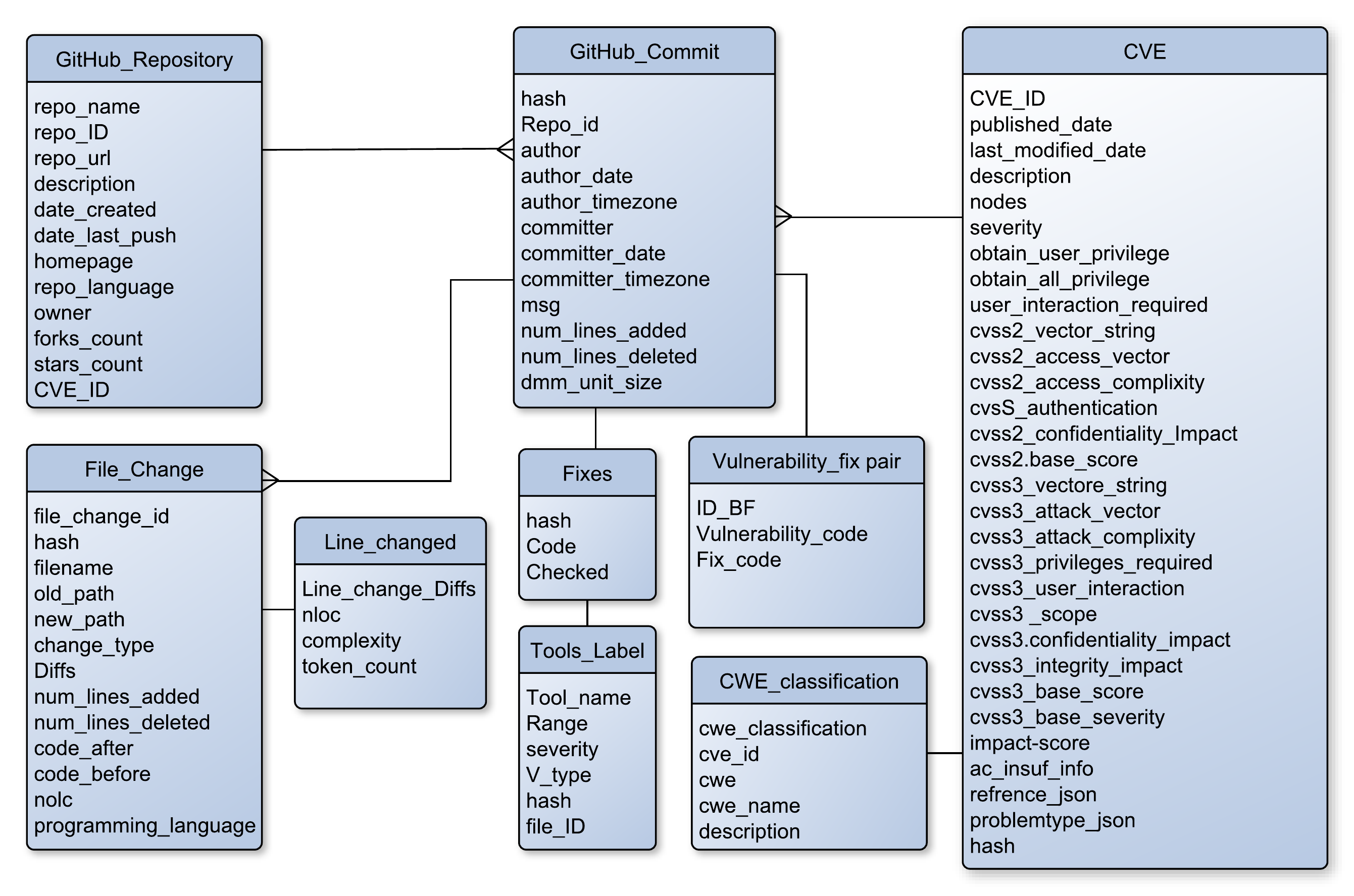}
\caption{AutoMESC dataset architecture}
\label{fig:classdiagram}
\end{figure*}   
This section describes the collected data and metadata from each data source. Furthermore, we explore the constructed dataset\footnote{\url{https://figshare.com/s/ea43905535cc1302267b}}.
AutoMESC primarily collects data and meta-data from the JSON vulnerability files from both NVD and projects on GitHub. Figure~\ref{fig:classdiagram} captures the overall data structure of our database, including the meta-data collected from NVD and GitHub (i.e., CVE vulnerabilities, OSS GitHub projects, CWE classification, commits, fixes, changed files, and changed lines). 

\subsection{Data Collected from CVE Records}
\label{cvssclass}
Each vulnerability has a CVE-ID that is a unique identifier. It also has a published\_date, description, user\_privilege, user\_interaction and last\_modified\_date. In addition, it has a severity ranking based on the Common Vulnerability Scoring System (CVSS) \cite{mell2007complete}. The NVD database provides two versions of the CVSS (i.e., CVSSv2 and CVSSv3).   
Figure \ref{fig:classdiagram} shows all the related data and meta-data attributes that are extracted from the CVE's JSON for each vulnerability in our dataset under the class ``CVE''. Each vulnerability is also linked with a unique hash that is linked with other classes in the dataset. It is directly linked with the commit class, as each CVE vulnerability has relevant source code repositories on GitHub and relevant commits that contain vulnerability source code and the corresponding fixes.  
In addition to labeling the vulnerabilities based on their CVSS, AutoMESC also classifies each vulnerability in the collected CVE according to CWE weakness types \cite{MITRE}. Therefore, AutoMESC collects the details of each CWE type associated with each collected CVE record. The collected meta-data of each CWE includes its name, description, and URL. Finally, each vulnerability in the CVE is associated with reference links to OSS repositories with commits that have the vulnerability and corresponding fixes. AutoMESC visits these links and provisionally clones the repositories to collect the related commits, vulnerability and corresponding fixes, and store them all in the database based on the unique hash of the commit listed in the CVE record.

\subsection{Data Collected from GitHub Repositories} Each repository on GitHub with vulnerabilities in Solidity or Vyper has a unique ID (i.e., repo\_ID) in our database. AutoMESC extracts the name, description, homepage, date of creations (i.e., date\_created), owner, the date of the last push (i.e, date\_last\_push) and other meta-data for each relevant OSS repository. Such meta-data can be used to focus on specific characteristics of relevant OSS repositories, such as the total number of vulnerabilities and corresponding fixes since the creation date of the repository, or the total number of vulnerabilities and corresponding fixes related to the popularity of the repository based on a minimum threshold on the number of forks. In order to avoid confusing Solidity repositories and Vyper repositories, AutoMESC also extracts the repository language (i.e, repo\_langugage). Finally, if the repository is associated with a CVE record then the CVE\_ID will contain the corresponding CVE identifier, otherwise it has a null value indicating that there is no CVE associated with the repository.  

\subsection{Commit Meta-data}
Commits in AutoMESC are the core of the database. AutoMESC collects commits from the collected OSS repositories for both Solidity and Vyper. Each commit has a unique hash and one commit is associated with a repository (repo\_id) and may be associated with a CVE record. For each commit, AutoMESC extracts the author, author\_date, author\_timezone, committer, committer\_date, committer\_timezone, and Delta Maintainability Model metrics (dmm\_unit\_size) for code changes \cite{di2019delta} meta-data. This meta-data can help, e.g., to analyze the time and date for each fix and other characteristics related to the author. AutoMESC also extracts the message (msg) for further details of the commit and the fix action (i.e., adding or deleting lines of the vulnerability code).

\subsection{Extracting Multiple Levels of Vulnerability-Fix Pairs} 
\begin{figure}[ht]
\centering
\includegraphics[width=.49\textwidth]{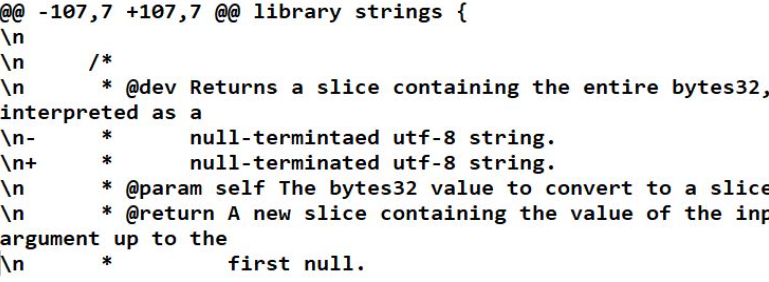}
\caption{Patch File Example: Code before, Code after and Code Diffs Strings}
\label{fig:diffs}
\end{figure}

AutoMESC extracts vulnerability-fix pairs at two different levels of granularity - one based on files and one on lines. AutoMESC associates each commit with one or more file\_changes, and each file change contains code diffs and the code of the file before (i.e., code\_before) and after the change (code\_after). Code diffs are presented in the same format delivered by Git. Figure \ref{fig:diffs} shows an example of the code before, code after and diffs. Any line starting with \textbackslash n- reflects the old code (code before) and lines starting with \textbackslash n+ indicate the new code (code after). Finally, diffs are listed in lines starting with @@ and ending with @@.
  
\subsection{Classification and Labeling of Extracted Vulnerabilities}
AutoMESC classifies and labels vulnerable code using various classifications to assist with ML feature extraction models and other ML models. First, as mentioned earlier, it extracts the relevant CWE type from the corresponding CVE record and annotates the vulnerability accordingly with the extracted CWE type. Then it extracts the description of each CWE from MITRE's\footnote{\url{https://cwe.mitre.org/data/index.html}} list of CWEs. Due to NVD's lack of distinction between CWE categories and CWE individual types, AutoMESC marks the field ``is\_category'' with true. Furthermore, AutoMESC classifies the extracted vulnerabilities based on the CVSS, as discussed before. Finally, AutoMESC analyzes the contract's code using 7 well-known tools, then classifies the file's vulnerability based on the results. It then labels the line numbers that contain the vulnerability with the vulnerability type name based on all the used tools, the severity of the vulnerability, and other meta-data.  
  

\subsection{Dataset Exploration}
Table \ref{table:statistics} provides a statistical overview of the first release of the AutoMESC dataset. The initial dataset contains 4.4K GitHub repositories, from which a total of 2.3K unique vulnerability-fix pair commits were extracted. 

\begin{table}[htbp]
\centering
\caption{Statistics overview of the AutoMESC dataset}
\label{table:statistics}
\footnotesize
\begin{tabular}{|c|c|c|c|c|} 
 \hline
 \# GitHub Repo. & \# Commits & \# Files & \# Contracts \\ 
\hline 
 4.4K & 2.3K & 6.7K & 10K \\
 \hline
  \# Methods & \# Lines& CVEs & \# of Vulnerabilities\\
\cline{1-4}
 6.8K& 5.241K& 0  & 97111\\
 \cline{1-4}

\end{tabular}
\end{table}
\begin{table}[htbp]
\centering
\caption{Severity levels distribution in AutoMESC data}
\label{table:severity}
\begin{tabular}{|c|c|c|c|} 
 \hline
 Severity  & 1: (Low) & 2:
(Medium) & 3: (High)\\
\hline  
Percentage& 85.3\% & 9.4\% & 5.2\%\\
 \hline
\end{tabular}
\end{table}
The total number of files in which there was a code change (that is, a vulnerability or a fix) is 6.7K. Moreover, a total of 10K contracts had vulnerability-fix changes. 

We also found out that CVEs have yet to be recorded for either Solidity or Vyper. According to this finding, it appears that the reported vulnerabilities on Ethereum smart contracts in the CVE records and NVD database exist in files other than (.vy) or (.sol) files. After manually analyzing the CVE records for Ethereum smart contracts, we found that most of the recorded vulnerabilities are interface-related. If future CVEs are recorded in Solidity or Vyper, AutoMESC will collect and store them in the database along with the related CWEs. 

There are around 97111 vulnerabilities in the first release of AutoMESC data. Among the most frequent vulnerabilities in AutoMESC data are hard-coded addresses, and implicit visibility levels, with more than 10K occurrences each. Table~\ref{table:vulstat} shows the most frequently occurring labeled smart contract vulnerabilities in AutoMESC data. Finally, Table \ref{table:severity} shows that most of the vulnerabilities in AutoMESC have low severity levels, with 5.2\% of the vulnerabilities having high severity.

\begin{table*}[htbp]
\centering
\caption{Top labeled smart contract vulnerability types using AutoMESC}
\label{table:vulstat}
\footnotesize
\begin{tabular}{|c|c|c|c|c|c|} 
 \hline
 Vulnerability Type & Total & Vulnerability Type & Total&Vulnerability Type & Total\\
\hline  
Hardcoded address&	11633& Implicit visibility level & 12536 &Unchecked low-level call & 3113\\
Compiler version not fixed&	7409 & Pure-functions should not read& 993 & Frozen Ether & 411 \\
Extra gas consumption&	3746 & Upgrade code to Solidity& 8431 & Pure-functions should not change state & 993\\
Costly loop	&4275& Multiplication after division& 627& Deprecated constructions& 519\\
Private modifier&	4694& Overpowered role& 1566  & Revert inside the if-operator & 936\\
Use of SafeMath	&1577& Comparison with block.timestamp& 5116  & View-function should not change state& 488\\
Revert inside the if-operator&	936 &  Replace multiple return values/struct & 4785 & Use of assembly& 2303\\
 \hline
\end{tabular}
\end{table*}
\section{Application scenarios}
\label{sec:app} 
In the following, we highlight a number of sample scenarios that illustrate how researchers and practitioners can use AutoMESC.
\subsection{Training Material} The AutoMESC dataset provides good examples of vulnerabilities and fixes that can be used as a learning resource for Ethereum smart contract developers. Using the provided vulnerability-fix pairs and classifications, they can learn how to resolve the vulnerabilities they face while implementing their smart contracts, and how to avoid the vulnerabilities in the future.  \\
\subsection{Vulnerability Localization and Prediction} Due to the automated approach, AutoMESC can assist future research in vulnerability detection as it continually collects newly posted vulnerabilities on OSS projects hosted on GitHub and disclosed Vulnerabilities on NVD. Additionally, the AutoMESC dataset is constructed using multiple levels of abstractions and meta-data about vulnerabilities. The dataset contains the actual vulnerability code that is pre-processed and labeled based on multiple tools published in the literature, which allows researchers to apply models directly to the vulnerable code. The provided meta-data can be used to improve the quality of models that use feature extraction, and can be used to train and test existing models. Several studies have already addressed automated vulnerability prediction and detection in Ethereum smart contracts, e.g., \cite{huang2022smart,lutz2021escort}, and using ML models. It is our belief that our dataset and framework will be a valuable asset to this line of research, and that our framework will provide the opportunity to evaluate the existing detection and prediction methods in a more comprehensive manner.

\subsection{Tool and Dataset Benchmark} With the constant updates of Solidity and Vyper, there will be a need for continuous data mining and constructing new datasets with the latest versions of both languages. Moreover, over time new vulnerabilities can occur, and some vulnerabilities will be deprecated, hence new analysis tools will be published. Therefore, AutoMESC and datasets constructed by it can be used for evaluating the correctness and other parameters of newly designed analysis tools for Ethereum smart contracts. Datasets with recent vulnerabilities can be collected and consolidated using AutoMESC. As a result, the constructed datasets can be used to evaluate the developed tools at a given point in time, and compare tools to each other in the form of a benchmark.

\subsection{Assist Empirical Studies on Smart Contract Vulnerabilities and Fixes}
In addition to providing several classifications, AutoMESC also provides meta-data per vulnerability, fix, repository, commit, owners, developers, CVE, CWE, and more to support large empirical studies. By examining such meta-data, empirical investigations can provide insight into how vulnerabilities are introduced to a smart contract, the characteristics of developers who implemented it, the patterns of vulnerabilities in Solidity and Vyper, the characteristics of vulnerability-related commits, common weaknesses in smart contracts, among other important metrics. Additionally, studying and analyzing some security patches can provide insight into some vulnerabilities and assist in detecting them early on.

\subsection{Program Repair for Vyper and Solidity Contracts}
AutoMESC constructs pairs of vulnerable code and its corresponding fixed code. The constructed pairs can be used in data-driven learning models for program repair. Automated smart contracts repair is still in its early stages \cite{yu2020smart}. As mentioned previously, the AutoMESC database has two levels of granularity. Therefore, it supports extracting particular fixes such as fixes with only one line, fixes with a specific number of lines, and fixes for the file. The findings of many studies such as \cite{morrison2015challenges} show that using line based granularity may improve the precision and recall of the ML model. Also, the training process can be focused on fewer code changes that are within the capacity of the used model or technology, which improves the predicted fix code.

\section{Evaluation}
\label{sec:results}
In this section, we evaluate the quality of the dataset and compare it against the related datasets discussed in Section~\ref{sec:survey}.

\begin{table*}[]
\centering
\caption{Characteristics of the datasets and data quality evaluation results}
\label{tab:evaluation}
\resizebox{\textwidth}{!}{%
\begin{tabular}{|cc|c|c|c|c|c|c|}
\hline
\multicolumn{2}{|c|}{Benchmark} &
  \cite{sujeet2022scrawld} &
  \cite{durieux2020empirical} &
  \cite{ren2021empirical} &
  \cite{zhang2020framework} &
  Gigahorse benchmarks &
  Our dataset \\ \hline
\multicolumn{1}{|c|}{\multirow{2}{*}{General characteristics}} &
  Purpose &
  Construct unbiased dataset &
  Evaluate SC tools &
  Construct unbiased dataset &
  Evaluate SC tools &
  Construct unbiased dataset &
  \begin{tabular}[c]{@{}c@{}}Construct unbiased dataset \\ for data-driven approaches\end{tabular} \\ \cline{2-8} 
\multicolumn{1}{|c|}{} &
  Vulnerabilities fixes &
  No &
  No &
  No &
  No &
  No &
  Yes \\ \hline
\multicolumn{1}{|c|}{\multirow{3}{*}{Amount of data}} &
  Data size &
  6.7K &
  143 &
  46K &
  266 &
  109 &
  6.7K \\ \cline{2-8} 
\multicolumn{1}{|c|}{} &
  Size of labeled data &
  6.7K &
  143 &
  564 &
  266 &
  109 &
  6.7K \\ \cline{2-8} 
\multicolumn{1}{|c|}{} &
  \# of attributes &
  6 &
  2 &
  2 &
  2 &
  7 &
  84 \\ \hline
\multicolumn{1}{|c|}{\multirow{5}{*}{Heterogeneity}} &
  Source of data &
  Etherscan &
  Etherscan &
  \begin{tabular}[c]{@{}c@{}}Etherscan, SolidiFI, CVE, \\ SWC and TCl\end{tabular} &
  \begin{tabular}[c]{@{}c@{}}Litrature papers and \\ existing datasets\end{tabular} &
  Etherscan &
  Github and CVE \\ \cline{2-8} 
\multicolumn{1}{|c|}{} &
  \# of SVT &
  8 &
  10 &
  42 &
  49 &
  9 &
  36 \\ \cline{2-8} 
\multicolumn{1}{|c|}{} &
  Supported languages &
  Solidity &
  Solidity &
  Solidity &
  Solidity &
  Solidity &
  Solidity and Vyper \\ \cline{2-8} 
\multicolumn{1}{|c|}{} &
  \# of used tools &
  5 &
  11 &
  9 &
  Manual &
  11 &
  7 \\ \cline{2-8} 
\multicolumn{1}{|c|}{} &
  CWE classification &
  No &
  No &
  Yes &
  No &
  No &
  Yes \\ \hline
\multicolumn{1}{|c|}{\multirow{3}{*}{Accuracy}} &
  Incompleteness &
  No &
  No evidence &
  No evidence &
  No evidence &
  0.92\% &
  No \\ \cline{2-8} 
\multicolumn{1}{|c|}{} &
  Redundancy &
  No &
  No evidence &
  No evidence &
  No evidence &
  No &
  No \\ \cline{2-8} 
\multicolumn{1}{|c|}{} &
  Inconsistency &
  No evidence &
  No evidence &
  No evidence &
  No evidence &
  No evidence &
  No evidence \\ \hline
\multicolumn{2}{|c|}{Timeliness} &
  No updates &
  No updates &
  No updates &
  No updates &
  No updates &
  Updates every 2 hour \\ \hline
\multicolumn{2}{|c|}{Accessibility} &
  Yes &
  Yes &
  Yes &
  Yes &
  Yes &
  Yes \\ \hline
\multicolumn{2}{|c|}{Trustworthiness} &
  Yes &
  Yes &
  Yes &
  Yes &
  Yes &
  Yes \\ \hline
\end{tabular}}
\end{table*}



To evaluate the dataset, we adopt the data quality taxonomy introduced by Bosu and Macdonell \cite{bosu2013taxonomy}, which rates data quality issues according to three dimensions: accuracy, relevance, and provenance. 
Accuracy refers to the correctness of the data and is measured based on the following metrics.
\begin{itemize}
    \item Incompleteness: missing data (md) (assigned as null, missing, or empty value). Incompleteness is measured by the percentage of missing data. 
    \item Redundancy: duplicate data (dd) that are exactly the same. This is calculated by the percentage of the duplicated data.
    \item Inconsistency: contradicting or non-matching data (cd). We measure Inconsistency by the percentage of contradicting data.
\end{itemize}
To calculate the Accuracy metrics ($Acc$), we use the equation \ref{eq:1}, where $n$ is the size of data, and $D$ can be md, dd, or cd:
\begin{align}
\label{eq:1}
  Acc=\frac{1}{n} \sum_{i=1}^{n} D_{i}  
\end{align}

The relevance class assesses the suitability of the data based on the following factors.
\begin{itemize}
    \item Heterogeneity: diversity of the data source.
    \item Amount of data: the size of the dataset, including the number of attributes.
    \item Timeliness: the age of the dataset, and how regularly it is updated.
\end{itemize}
Provenance refers to the origin of the dataset. It is evaluated based on two metrics.
\begin{itemize}
    \item Accessibility: data available to the public.
    \item Trustworthiness: the collection of data is documented and can be replicated.
\end{itemize}

Table~\ref{tab:evaluation} summarizes the general characteristics of the datasets and the quality evaluation results.

\subsection{Accuracy} For this class, we consider the datasets with an acceptable format such as CSV, JSON, etc. The datasets \cite{durieux2020empirical,ren2021empirical,zhang2020framework} are not evaluated for this class, as they are folder-based datasets, where the smart contracts are grouped in folders based on their vulnerabilities. These type of datasets cannot be used in data-driven approaches as the majority of them has only two attributes. Hence, in this class, we will only consider the Gigahorse benchmarks\footnote{\url{https://github.com/nevillegrech/gigahorse-benchmarks}} and the Sujeet Yashavant et al.~\cite{sujeet2022scrawld} dataset.

In terms of incompleteness, the Gigahorse benchmarks showed 0.92\% missing data, while our dataset and Sujeet Yashavant et al. did not have any missing data. 
No redundancy was found in any of the datasets, even in the folder-based datasets. In all datasets, there was no evidence of any inconsistency issues. However, it was challenging to determine inconsistency in the datasets, as most of the datasets used different tools to detect vulnerabilities, and some of these tools might not have the same conclusion.

\subsection{Relevance} In terms of heterogeneity, we have analyzed the source of data, the number of tools used to label the data, supported vulnerability types, supported languages, and if CWE classification is supported. All datasets are heterogeneous, where (i) the data were collected from at least two data sources, or (ii) they support multiple vulnerabilities. However, all the existing datasets only support the Solidity programming language, and not all of them support CWE classification. 

The amount of data is determined based on the size of the dataset, the size of labeled data, and the number of attributes for each dataset. Some of these datasets are in the format of folders, where the attributes are the vulnerability type and the code file. 

In terms of timeliness, the existing datasets are not updated after the data are made public. In our case, AutoMESC is designed so that the dataset can be updated every two hours. 

\subsection{Provenance} All the analyzed datasets are available publicly, and the collection of data is documented in detail as research papers or in GitHub instructions. 


\subsection{Evaluation Summary}
Based on the above comparison, the AutoMESC dataset provides heterogeneous data with a variety of attributes that can be adapted for different data-driven research projects in the area of smart contract vulnerabilities. The dataset overcomes the current limitations in dataset timeliness where it updates the dataset every two hours based on newly disclosed vulnerabilities. This improves the quality of the AutoMESC dataset since smart contracts evolve rapidly, where new vulnerabilities are discovered and several vulnerability types become deprecated over time.

\section{Threats to Validity}
\label{sec:threats}
\subsection{Internal Validity} A possible internal validity threat may result from an implementation bug in the codebase due to the complexity of the AutoMESC framework. We thoroughly tested the framework to address this concern. Moreover, we made the framework and the dataset publicly available so researchers and developers can verify the framework. 


Each tool selected for vulnerability detection is neither sound nor complete. In some cases, a tool may generate a high number of false positives or false negatives. To mitigate this, we assessed whether a vulnerability exists or not based on the majority approach. However, such an approach could fail if the majority of the tools result in false positives or false negatives. These issues can be addressed either by integrating more tools or by manually verifying vulnerabilities through crowdsourcing or other methods. Nevertheless, these approaches require considerable time and resources, and we can apply them incrementally.

\subsection{External Validity} A threat to external validity is the generalizability of AutoMESC. We built our dataset using publicly disclosed vulnerabilities and their fixes. These pairs may not reflect all vulnerabilities and fixes in smart contracts, particularly those not reported. To reduce this threat, we also collected vulnerability information from CVE and checked if these vulnerabilities are linked with Github repositories. If there is no match between collected vulnerabilities from CVE and Github repositories, we include these vulnerabilities in our dataset.  

\section{Conclusion}
\label{sec:conclusion}
Several traditional and data-driven methods and tools were proposed in literature for improving the security of smart contracts, detecting their vulnerabilities, and perhaps fixing them. Nevertheless, data-driven research on smart contracts vulnerabilities and fixes is still in its infancy, and a comprehensive dataset with smart contract's vulnerabilities and corresponding fixes is missing to support such research. This paper proposes a method of automatically mining and classifying smart contract vulnerabilities and their fixes under a fully automated framework called AutoMESC. AutoMESC constructs a dataset of vulnerabilities and fixes for smart contracts written in the most popular smart contract languages (Solidity and Vyper) mined from OSS projects hosted on GitHub and CVE records. The constructed dataset is enriched with meta-data that can support machine learning models for feature extraction. The initial dataset consists of approximately 6.7k smart contract vulnerability-fix code pairs with various levels of granularity and meta-data. In addition to opening up new opportunities for researchers in the smart contract research and empirical software engineering research, our framework and dataset can also be employed to identify smart contracts vulnerabilities, classify them, predict severity levels, automate their repair and many more. In the future, we intend to empirically investigate the relationship between smart contract's vulnerabilities and their fixes, and the patterns they show at different abstraction levels. Additionally, we plan to use deep learning models for AutoMESC's classification process to predict the type of newly discovered vulnerabilities.

\section*{Acknowledgment}
This work was supported by blinded.

\bibliographystyle{ieeetr}
\bibliography{sample-base}

\end{document}